\documentclass{Rinton-P9x6}
\usepackage{amsfonts}
\usepackage{amsmath}

\def\cH{{\mathcal H}}

\def\C{{\mathbb C}}

\newcommand\trnorm[1]{\left\| #1 \right\|_1}
\newcommand\hsnorm[1]{\left\| #1 \right\|_2}

\newcommand\ket[1]{| #1\rangle}

\def\tr{\operatorname{tr}}
\def\id{\operatorname{id}}

\def\idty{{\leavevmode{\rm 1\mkern -5.4mu I}}}

\begin{document}

\title{Entropy-Energy Balance\\
in Noisy Quantum Computers}

\author{Maxim Raginsky}

\address{Center for Photonic Communication and Computing, ECE Department\\
Northwestern University, 2145 N. Sheridan Rd., Evanston, IL 60208-3118, USA\\
E-mail: maxim@ece.northwestern.edu}

\maketitle

\abstracts{
We use entropy-energy
arguments to
assess the limitations on the running time and on the system size, as
measured in qubits, of noisy macroscopic circuit-based quantum computers.}

Quantum information is essentially a nonequilibrium theory in the
sense that its basic primitives are decomposed into temporal sequences
of local operations that are performed on physical systems driven far
from equilibrium.  The requirement of locality becomes all the more
important when we consider quantum information processing in
macroscopic systems (multiqubit quantum computers), because local
operations will be used to model both the computation proper and the
errors affecting it.

An important issue to address is the stability of a macroscopic
quantum computer (MQC) in the presence of noise.  There are two notions of
stability for quantum-mechanical states,\cite{sew} namely {\it global} and
{\it local thermodynamic stability}.  For a microscopic system (i.e.,
one with
finitely many degrees of freedom) these are equivalent and
amount to the following.  Let $H$ be the system Hamiltonian, $\beta$
the inverse temperature. Then a state (density operator)
$\rho$ is thermodynamically stable (both globally and locally) if it
minimizes the free energy functional $F_\beta(\rho) = \tr{(\rho H)} -
(1/\beta)S(\rho)$, where $S(\rho) = -\tr{(\rho \ln \rho)}$ is the von
Neumann entropy.  However, the two stability notions are rather
different for systems with infinitely many degrees
of freedom. A state $\rho$ is globally thermodynamically
stable (GTS) if it minimizes the specific free energy (i.e., free
energy ``per particle''), and locally
thermodynamically stable (LTS) if no {\it local} modification of it yields a state with lower specific free energy.  It can be
shown that any GTS state is also LTS, but the converse is generally false.

States that are LTS but not GTS are referred to as {\it metastable
  states}.\cite{sew}  An apt example comes from laser physics.
  Imagine an active medium consisting of a large number of three-level
  atoms with the states $\ket{0},\ket{1},\ket{2}$, each atom initially in
  the ground state $\ket{0}$.  The atoms are pumped to the level
  $\ket{2}$, and then decay nonradiatively to $\ket{1}$.  This results
  in a relatively long-lived metastable state of population
  inversion.  As far as MQC's are concerned, we are interested mainly
  in their metastable states, and the goal of error correction (and
  of judicious MQC design in general) is to preserve metastability for
  the duration of the computation.  Specifically, we can isolate two separate aspects of
  (meta)stability for MQC's --- temporal and spatial. The former
  refers to the maximum number of operations that can be carried out
  on a noisy MQC using given energy resources, while the latter is
  related to the maximum number of qubits that can be processed
  reliably on a noisy MQC.  We will address these two issues using
  entropy-energy arguments,\cite{ss} a standard technique in
  statistical physics.  Due to lack of space, our presentation will be
  rather sketchy; we will supply the details in a separate
  paper.\cite{rag2} We note that even though we consider {\em
  macroscopic systems}, there is no need to pass to the thermodynamic
  limit because we are concerned with {\em local} thermodynamic stability.

First of all, let us fix a model of an $N$-qubit quantum computer. We may
imagine its operation as a sequence of alternating computation and
noise steps.  For a circuit-based QC, a computation step is an
application of a tensor product of one- and two-qubit universal
quantum gates, while a noise step is described by a general
trace-preserving completely positive linear map $T$ on density matrices.
We will henceforth refer to such maps as channels (note that any
unitary transformation is an instance of a channel).
Specifically, if $\cH \simeq \C^2$ is a single-qubit Hilbert space,
then the state of the $N$-qubit computer is a density
operator $\rho$ on $\cH^{\otimes N}$.  The composition of a
computation step and a noise step is a mapping $\rho \mapsto T(U\rho
U^*)$, where $U$ is unitary. We adopt the model of local stochastic noise,\cite{aha}
so that $T$ can be written in the form $(1-\epsilon)\id + \epsilon
R^{\otimes N}$, where $\id$ is the identity channel, $R$ is a channel
on $2\times 2$ density matrices, and $\epsilon$ is a small positive
number that quantifies the noise strength.

We make two simplifying assumptions.  The first one is needed for
the analysis of the temporal stability and concerns the
noisy channel $R$.  Namely, we will take $R$ to be bistochastic, i.e.,
$R(\idty) = \idty$, and strictly contractive,\cite{rag} which means
that there exists a constant $C \in [0,1)$ (called the {\it
    contraction rate}) such that, for any two
  density operators $\rho,\sigma$, $\trnorm{R(\rho)-R(\sigma)} \le
  C\trnorm{\rho-\sigma}$, where $\trnorm{X} = \tr |X|$ is the trace norm. For instance, the much studied
  depolarizing channel, $R(\rho) = (1-\lambda)\rho + \lambda \idty/2$ for some
  $\lambda \in [0,1)$, fits this description.  Our second assumption,
    necessary only for the analysis of the spatial stability,
    states that the number of gates applied during any computation
    step is bounded by a positive number $K$ that depends only on the
    particular algorithm, but not on the number of qubits $n$.  This
    means that our analysis will be inapplicable to highly
    parallelized computation.\cite{mn}

We will rely upon the following two theorems, the proofs of
which are given elsewhere.\cite{rag2,rag1}\\

\noindent{{\bf Theorem 1.} Consider the channel $R_N = R^{\otimes N}$,
  where $R$ is a bistochastic strictly contractive channel on $2\times
  2$ density matrices with contraction rate $C$.  Then, for any $2^N
  \times 2^N$ density matrix $\rho$ and any positive integer $m$, we
  have $S[R^m_N(\rho)] - S(\rho) \ge \frac{1-C^{2m}}{2}\hsnorm{\rho -
  2^{-N}\idty}^2$, where $S(\rho) = -\tr{(\rho \ln \rho)}$ is the von
  Neumann entropy and $\hsnorm{X} = \sqrt{\tr(X^*X)}$ is the
  Hilbert-Schmidt norm of the operator $X$.}\\

\noindent{{\bf Theorem 2.} Let $\rho_i$, $i=1,\ldots,n$, be $n$
  mutually commuting density operators.  Suppose that there exists a
  constant $\kappa \ge 0$ such that, for any $i$, $\sum_{j \neq
  i}\tr{(\rho_j \rho_i)} \le \kappa$.  Let $\rho =
  n^{-1}\sum^n_{i=1}\rho_i$. Then $S(\rho) \ge
  n^{-1}\sum^n_{i=1}S(\rho_i) + \ln n - 2\sqrt{\kappa}$.\\

Let us first consider temporal stability.  Suppose that our computer
operates on $N$ qubits, and that we are given ``energy resources'' $E$
(this parameter could be determined, e.g., from the so-called ``control Hamiltonian''
representation of quantum computation\cite{ogkl}).  We are interested
in the entropy increase after $m$ steps of noisy computation.
Since entropy can only be produced by the channel $T$ and not
by the unitarily implemented gates, the entropy gain is given by
$\Delta S(\rho,m) \equiv S[T^m(\rho)] - S(\rho)$, where $\rho$ is the
initial state of the computer.  Now $T^m = \sum^m_{j=0} {m \choose j}
(1-\epsilon)^{m-j}\epsilon^j R^j_N$, where $R_N = R^{\otimes N}$, so
that by concavity of the von Neumann entropy we have
$\Delta S(\rho,m) \ge \sum^m_{j=0} {m \choose
  j}(1-\epsilon)^{m-j}\epsilon^j \left\{S[R^j_N(\rho)]-S(\rho)
\right\}$.  Using Theorem 1 we can write
\begin{equation}
\Delta S(\rho,m) \ge \zeta_{\rho,N}\sum^m_{j=0}{m \choose j}
(1-\epsilon)^{m-j}\epsilon^j (1-C^{2j}),
\label{eq:step1}
\end{equation}
where $\zeta_{\rho,N} \in [0, (2^N-1)/2^{N+1}]$ depends only on
$\rho$ and $N$.  Carrying out the summation in (\ref{eq:step1}) we obtain
\begin{equation}
\Delta S(\rho,m) \ge
\zeta_{\rho,N}\left\{1-[1-\epsilon(1-C^2)]^m\right\} =
\zeta_{\rho,N}m\epsilon(1-C^2) + o(\epsilon),
\label{eq:step2}
\end{equation}
where $o(\epsilon)$ stands for terms that go to zero faster
than $\epsilon$ as $\epsilon \rightarrow 0$ and therefore can be neglected
when $\epsilon$ is sufficiently small.  The corresponding free-energy shift is $\Delta
F_\beta(\rho,m) \equiv E-(1/\beta)\Delta S(\rho,m)$, where $\beta$
is the inverse temperature.  It is now easy to see from the estimate
(\ref{eq:step2}) that $\Delta F_\beta(\rho,m)$ will be negative for
all $m \ge \bar{\zeta}_{\rho,N,T}\beta E$, where
$\bar{\zeta}_{\rho,N,T}$ is a constant that depends on $\rho$, $N$,
and $T$.  In other words, the state of the MQC will not be metastable
for large enough $m$, or, to put it bluntly, the MQC will ``crash''
after $O(\beta E)$ operations.  However, because the produced entropy
may be at least partially drained off by error correction, this
estimate merely tells us how often it should be carried out.

We pass now to spatial stability. We note that in a circuit-based
computer each computational step is an application of a bounded
number of universal quantum gates, and the initial state of the
computer may be taken separable without loss of generality.
Therefore at every instant of the computation the set $\{
1,\ldots,N\}$ can be partitioned into $L$ disjoint clusters
$C_1,\ldots,C_L$, so that the state of the computer has the form
$\rho = \bigotimes^L_{l=1}\rho^{(l)}$, where $\rho^{(l)}$ is the state of the
qubits in the $l^{\rm th}$ cluster.\cite{aha}  Let us suppose for
simplicity that the clusters all have the
same size $d$, so that the computer operates on $N=Ld$ qubits.  Let
us partition the set $\{C_1,\ldots,C_L\}$ into $n$ disjoint blocks
$S_1,\ldots,S_n$, each of which contains $k$ clusters; thus
$N=nkd$.  Now, for each $i = 1,\ldots,n$, define $T_i$ to be the
channel that acts as a depolarizing channel $\sigma \mapsto
(1-\lambda)\sigma + \lambda \idty/2^d$ on the state of each of the clusters in
$S_i$ and leaves the remaining clusters intact.  Then $\rho_i \equiv
T_i(\rho)$ are $n$ mutually commuting density operators for which we
have the elementary estimate
\begin{equation}
\tr \sum_{j \neq i} \rho_j \rho_i \le (n-1)\left(1-\lambda +
\frac{\lambda}{2^d}\right)^{2k}.
\label{eq:trest}
\end{equation}
Note that for fixed $n$, $\lambda$, and $d$ we can always choose $k$
so large that the r.h.s. of (\ref{eq:trest}) is bounded above by unity.  Thus the conditions of
Theorem 2 are satisfied with $\kappa = 1$, and the entropy gain due to the channel
$T \equiv n^{-1}\sum^n_{i=1}T_i$ is $O(\ln n)$.  Because the energy
shift due to $T$ depends only on the number $K$ of gates applied
during each computational step, the free-energy shift will be negative
for large enough $n$.  We may thus conclude that the number of qubits
that can be processed reliably (that is, kept in a metastable state)
in a MQC is bounded from above as $O(e^{\beta E})$, where $\beta$ is
the inverse temperature, and $E$ depends on $K$.

To summarize, we have shown that there exist upper
bounds on the running time and on the size (in qubits) of
circuit-based noisy MQC's.  However, these constraints become significant
only when $\beta E$ is small (as in, e.g., ensemble quantum
computation using NMR).  Furthermore, our analysis applies only to
circuit-based computers without parallelization; massively
parallel non-circuit models, such as the ``one-way quantum computer''
of Briegel and Raussendorf,\cite{br} are likely to be intrinsically
thermodynamically stable.

\section*{Acknowledgments}
This work was supported by the U.S. Army
Research Office and by the Defense Advanced Research Projects Agency.

\end{document}